\documentclass[reprint,amsmath,amssymb,aps,pre,showpacs]{revtex4-1}
\usepackage{graphicx}
\usepackage{dcolumn}
\usepackage{bm}
\usepackage{hyperref}
\usepackage{array}
\usepackage{subfigure}

\begin{document}

\title{Synchronized interval in random networks: The role of number of edges}
\author{Suman Acharyya}
\email{suman.acharyya@icts.res.in}
\affiliation{International Centre for Theoretical Sciences (ICTS), Survey No. 151, Shivakote, Hesaraghatta Hobli, Bengaluru North - 560 089, India.}


\date{\today}

\begin{abstract}
We study the synchronized interval in undirected and unweighted random networks of coupled oscillators as a function of the number of edges. In many coupled oscillator systems, synchronization is stable in a finite interval of coupling parameter, which we define as the synchronized interval. We find in random networks, the width of the synchronized interval is maximum for an optimal number of edges. We derive analytically an estimation for this optimal value of the number of edges. In small networks, the analytical estimation deviates from the numerical results. However, in large networks the analytical and numerical results are in excellent agreement.
\end{abstract}

\maketitle

Complex networks have become standard tool to model complex systems of interacting elements, where the \emph{nodes} or \emph{vertices} of the network represent the elements and the \emph{links} or \emph{edges} represent the interactions~\cite{Newman-book,Dorogovtsev_book2003,NewmanSIAM2003,Boccaletti2006,AlbertRMP2002}. 
The interacting elements on complex networks are capable of exhibiting rich collective dynamics such as synchronization~\cite{KurthsBook}, chimera states~\cite{AbramsPRL2004}, amplitude deaths~\cite{Mirollo1990} and multistability~\cite{FeudelIJBC2008}. Synchronization which emerges as one of the fundamental collective dynamics of the interacting dynamical units~\cite{BoccalettiPhysRep2002,Strogatz-book,Winfree-book,KurthsBook,Winfree1967,Arenas2008}, is ubiquitous in many natural and man made systems. These include the synchronized flashing of fireflies~\cite{BuckScience1968}, the crowd synchrony on the London Millennium bridge~\cite{StrogatzNature2005}, synchronous firing of neurons~\cite{Singer2008}, synchronization in an array of Josephson junction~\cite{VlasovPRE2013}. Thus, synchronization has widespread applications in many  areas at the interface of physics with biology and engineering as well as the social sciences~\cite{KurthsBook,KocarevPRL1995}.

Depending on their structural properties, complex networks are categorized in different classes. Amongst them the most studied are regular networks or lattices, random networks~\cite{Erdos1960}, small-world networks~\cite{Watts1998} and scale-free networks~\cite{Barabasi1999}. 

The structure of a complex network has a nontrivial impact on the synchronization dynamics of coupled oscillators on it~\cite{Arenas2008,PecoraPRL1998,BarahonaPRL2002,NishikawaPRL2003}. The small-world and scale-free networks synchronize far more easily than, e.g. random networks~\cite{BarahonaPRL2002}. However, for scale-free networks the inhomogeneity in the degree distribution suppresses synchronization~\cite{NishikawaPRL2003}

Indeed, the search for network structures with better synchronization properties has been a subject of intense research in recent years~\cite{MotterNatPhys2013,NishikawaPhysicaD2006,DonettiPRL2005,JaliliIEEETNNLS2013}. In particular, such studies have focused on enhancing synchronizability in random complex networks by adding suitable weights to its edges~\cite{MotterEPL2005,MotterPRE2005,MotterAIP2005,NishikawaPRE2006}, by rewiring its edges~\cite{DonettiPRL2005,DonettiJSM2006,DonettiJPA2008,JaliliIEEETNNLS2013} or by suitable insertion of a certain number of additional edges~\cite{SchultzPRE2016}. While the above mentioned methods are  successful in enhancing synchronizability, they have well-known drawbacks when applied to many real-world systems~\cite{PadeSciRep2015,TimmeNJP2012}. It is therefore essential to look for alternate strategies to enhance synchronizability in complex networks. 

In this article, we focus on random networks. With extensive numerical simulations and analysis we found that there is a global optimal value of the number of edges for which the chance of a generated random network to remain synchronizable in the widest interval of coupling parameter is higher. Ref.~\cite{NishikawaPNAS2010} revealed that deleting edges and/or adding negative interaction some times can be beneficial to synchronization, and provided the quantization number of total connections which maximizes synchronizability in networks. Ref.~\cite{NishikawaPhysicaD2006} studied the optimal structures for maximal synchronization with minimal number of edges. These studies considered the ratio of the largest and smallest nonzero eigenvalues of the Laplacian matrix of networks as their target function. In this article, we consider the width of the synchronized interval as our target function~\cite{AcharyyaEPL2012,AcharyyaPRE2015}. Starting with a complete network, we remove edges randomly and study the change in the width of synchronized interval as a function of the number of edges. Our numerical experiments reveal the existence of global optimal value for the number of edges for which the width of synchronized interval is maximum. We support the numerical observations with analytic derivation. We note here that the purpose of the present article is not to find the optimal structure for network synchronization. Instead, we search for the optimal value of the number of edges for which a random network has high chances to be synchronizable in wide range of coupling parameter. Our results assume special importance in many physical networks those are prone to removing and inserting edges, for example power grid networks~\cite{RohdenPRL2012,WitthautPRL2016} show need of addition or removal of connection lines to meet the daily demand of power consumptions, in transportation networks new roads are build to tackle the increasing traffic. Random addition of edges can lead to functional failure in networks~\cite{Braess2005,TimmeNJP2012}. Thus, it is important to study what will be the optimal number of edges for which a network performs best.

\begin{figure}[ht]
\includegraphics[width=\columnwidth]{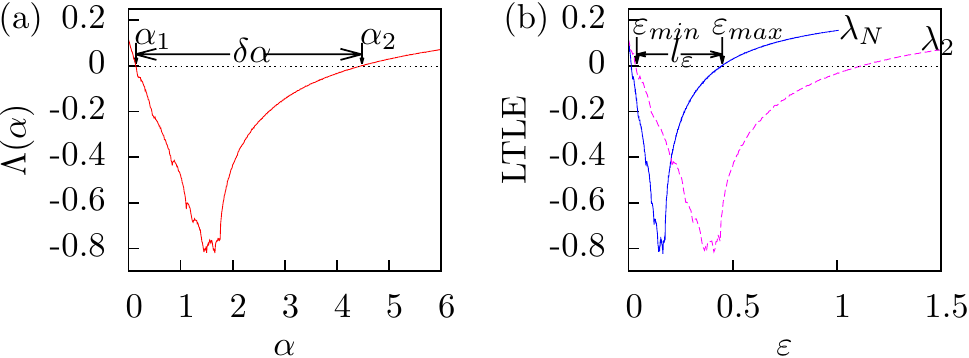}
\caption{\label{MSFRossler}(color online) (a) The Master Stability Function (MSF) $\Lambda(\alpha)$ for $x$-coupled R\"ossler oscillator is plotted as a function of the parameter $\alpha$. The MSF is negative in interval ($\alpha_1,\alpha_2$), here $\alpha_1\approx0.14$ and $\alpha_2\approx4.48$.  
(b)The two largest transverse Lyapunov exponents (LTLEs), corresponding to the maximum (blue solid line) and the minimum (pink dotted line) nonzero eigenvalues of the coupling matrix $L$ are plotted as functions of coupling parameter $\varepsilon$. In the interval $(\varepsilon_{min},\varepsilon_{max})$ all transverse Lyapunov exponents are negative and synchronization is stable. We define this interval of coupling parameter as the \emph{synchronized interval} $(l_{\varepsilon} = \varepsilon_{max} - \varepsilon_{min})$. The R\"ossler parameters are $a=0.2$, $b=0.2$ and $c=7.0$.}
\end{figure}

We consider connected, undirected and unweighted network of $N$ nodes and $E$ edges. Identical oscillators are placed at the nodes of the network. The dynamics of the $i$-th oscillator is given as
\begin{equation}
\dot{x}_i = F(x_i) - \varepsilon \sum_{j=1}^N L_{ij}H(x_j); \; i = 1,\ldots, N
\label{i_dyn}
\end{equation}
where, $x_i \in R^d$ is the $d$ dimensional state variable of the oscillator $i$ and $F: R^d \mapsto R^d$ provides the dynamics of individual oscillator, $H: R^d \mapsto R^d$ is the coupling function and $\varepsilon$ is overall coupling parameter.
The coupling matrix is $L=[L_{ij}]$ and it is defined as $L_{ij}=-1;\; i\neq j$ if nodes $i$ and $j$ are coupled, otherwise $L_{ij}=0;\; i\neq j$, and the diagonal entries are $L_{ii}=-\sum_{j\neq i}L_{ij}$. The row sum of the coupling matrix $L$ satisfies $\sum_{j=1}^N L_{ij}=0$ which guarantees synchronization solution is invariant. The coupling matrix $L$ is symmetric and all its eigenvalues are real and nonnegative. The smallest eigenvalue of $L$ is always zero and the corresponding eigenvector is $(1,1,\ldots,1)^T$. We arrange the eigenvalues of $L$ in ascending order $0=\lambda_1<\lambda_2\leq\ldots\leq\lambda_N$.
The multiplicity of the zero eigenvalue of $L$ is equal to the number of disconnected components in the network. Since we are interested in the synchronization on networks, we consider the networks are always connected and the second smallest eigenvalue of $L$ is always greater than zero $(\lambda_2>0)$.

For suitable coupling parameter $\varepsilon$ and coupling function $h$ the coupled oscillators undergo complete synchronization to an invariant solution $x_1 = x_2 = \ldots = x_N = s$ where $s$ is the solution of an isolated dynamics $\dot{s} = f(s)$. As a consequence of complete synchronization, the motion of the coupled oscillators collapse to a $d$-dimensional subspace of the phase space, which is known as synchronization manifold~\cite{PecoraChaos1997}. The stability of the complete synchronization against small perturbations in the direction transverse to the synchronization manifold are analyzed by determining the Lyapunov spectrum calculated at the synchronous
solution~\cite{PecoraPRL1990,HeagyPRE1994,HeagyPRER1995,HeagyPRL1994,Heagy.PhysRevLett.74.4185.1994}.
Pecora and Carroll~\cite{PecoraPRL1998} developed an elegant tool for determining stability of complete synchronization of coupled identical oscillators on networks, namely the Master Stability Function (MSF) which is the largest Lyapunov exponent calculated from a set of equations known as the Master Stability Equations. The MSF at once resolve the stability problem for all networks with equivalent coupling function. The linearized dynamics of the small perturbations can be written as  $\dot{\eta}_k = [DF(s) - \alpha_k DH(s) ] \eta_k; k= 1,\ldots, N$, where $\eta_k$ is the small perturbation in the direction of the $k$-th eigenvector of coupling matrix $L$ and  $\alpha_k=\varepsilon\lambda_k$. $DF(s)$ and $DH(s)$ are the Jacobian matrices of the dynamics $F$ and the coupling function $H$ evaluated at the synchronized solution $s$. The linearized equations can be put in the form the MSE by introducing a generic parameter $\alpha$ for all $\alpha_k$. Thus, the MSE is 
\begin{equation}
\dot{\eta} = [Df(s) - \alpha Dh(s)] \eta.
\label{mse}
\end{equation}
The MSF $\Lambda$ is the largest Lyapunov exponent calculated from the MSE Eq.~(\ref{mse}) as a function of $\alpha$. Synchronization is stable in the region where the MSF is negative. In Fig.~(\ref{MSFRossler})a, the MSF for $x$-component coupled chaotic R\"ossler oscillators~\cite{Rossler} is plotted as a function of $\alpha$. The dynamics of the R\"ossler oscillator is given by $\dot{x}=-y-z,\dot{y}=x+ay,\dot{z}=b+z(x-c)$~\cite{Rossler}, where $a=b=0.2$, $c=7.0$. The MSF is negative in a finite interval of $\alpha$, let the interval be $(\alpha_1,\alpha_2)$ (Figure~(\ref{MSFRossler} a)). In literature, these MSFs are known as Type-II MSF~\cite{HuangPRE2009}. The perturbations corresponding to the smallest eigenvalue $\lambda_1=0$ of coupling matrix $L$ is parallel to the synchronization manifold and does not affect the stability of the synchronization solution. Thus, the synchronization is stable when $\varepsilon\lambda_i;\; i=2, \ldots, N$ fall in the negative region of the MSF, i.e. $(\alpha_1 < \varepsilon\lambda_2 \leq \ldots \leq \varepsilon\lambda_N < \alpha_2)$. Thus the condition for stable synchronization is
\begin{equation}
\frac{\lambda_N}{\lambda_2} < \frac{\alpha_2}{\alpha_1}.
\label{sync_cond}
\end{equation}
The LHS of Eq~(\ref{sync_cond}) depends on the network topologies and the RHS depends on the dynamics of the coupled oscillators, coupling function and synchronization solution. A network is more synchronizable when its eigenvalue ratio is less.

In Fig.~(\ref{MSFRossler})b, display a typical stability diagram of a network where we plotted the two largest transverse Lyapunov exponents (LTLEs) corresponding to the maximum (solid blue curve) and minimum (dashed pink curve) nonzero eigenvalue of the coupling matrix $L$ respectively. 

As the coupling parameter $\varepsilon$ is increased from zero, all coupled oscillators on the network synchronize when the coupling parameter exceeds a critical value $\varepsilon=\varepsilon_{min}=\alpha_1/\lambda_2$ (at this point the LTLE corresponding to the smallest nonzero eigenvalue $\lambda_2$ become negative) and remain synchronized until $\varepsilon=\varepsilon_{max}=\alpha_2/\lambda_N$ (the LTLE corresponding to the largest eigenvalue $\lambda_N$ become positive at this point) beyond which the coupled oscillators desynchronize through short-wavelength bifurcation~\cite{Heagy.PhysRevLett.74.4185.1994,Acharyya2011}. In the interval $(\varepsilon_{min},\varepsilon_{max})$ the Lyapunov exponents corresponding to the remaining nonzero eigenvalues of coupling matrix $L$ are negative and synchronization is stable. We define this interval of coupling parameter as the \emph{synchronized interval} and its width is given as  $l_{\varepsilon}  = \varepsilon_{max} - \varepsilon_{min} = {\alpha_2}/{\lambda_N} - {\alpha_1}/{\lambda_2}$. We express this width in a more convenient form
\begin{equation}
l_{\varepsilon} = \frac{\delta\alpha}{\lambda_N} - \alpha_1\left[ \frac{1}{\lambda_2} - \frac{1}{\lambda_N} \right]
\label{sync_interval2}
\end{equation}
where, $\delta\alpha=\alpha_2-\alpha_1>0$. Unlike the eigenvalue ratio $\lambda_N/\lambda_2$ which is independent of the dynamics, the width of the synchronized interval $l_{\varepsilon}$ of a network depends on the dynamics of the coupled oscillators.
For networks with all equal nonzero eigenvalues $\lambda_2=\ldots=\lambda_N$, the synchronized interval is $\delta\alpha/\lambda_N$, and the eigenvalue ratio is also minimum $(\lambda_N/\lambda_2=1)$. Complete networks, outward star networks belong to such class of networks~\cite{NishikawaPhysicaD2006}.

\begin{figure}
\includegraphics[width=\columnwidth]{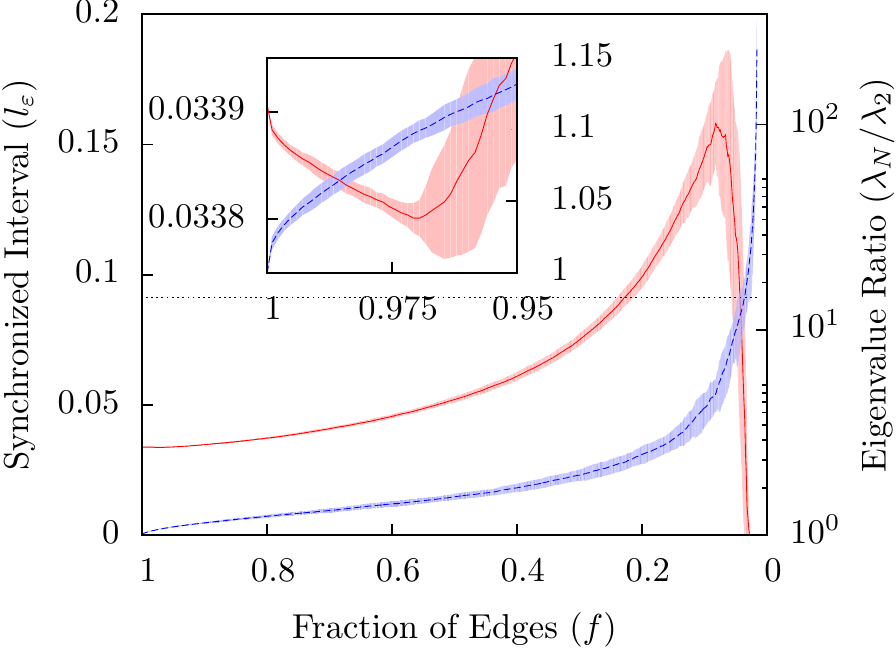}
\caption{\label{lepsilon_eigrat_128_stddev}(color online) The width of the synchronized interval $l_{\varepsilon}$ (solid red curve, left axis) for $x$-component coupled R\"ossler oscillators and the eigenvalue ratio $\lambda_N/\lambda_2$ (dashed blue curve, right axis) are plotted as functions of the fraction of existing edges to the complete network $f=2E/N(N-1)$; $N$ and $E$ are the number of nodes and existing edges respectively, here $N=128$ nodes. The initial network is undirected, unweighted and completely connected, i.e. $f=1$, $\lambda_N/\lambda_2=1$, and $l_{\varepsilon}=\delta\alpha/N\approx0.034$. Edges are removed randomly from the complete network. The width of the synchronized interval $l_{\varepsilon}$ increases  as edges are removed randomly from the complete network, and for an optimal value of the fraction of edges $f=f^{\rm{optimal}}\approx 0.082$, the synchronized interval achieves maximum value $l_{\varepsilon}=l_{\varepsilon}^{\rm{maximum}}\approx 0.158$. The $l_{\varepsilon}$ decreases to zero as further edges are removed. In the beginning, when few edges are removed randomly from the complete network, the width of synchronized interval first decreases and then increases when more edges are removed (see inset). This is an averaged result on 100 realizations. The shaded regions are the standard deviations of the accompanying curves. The R\"ossler parameters are same as in Fig.~(\ref{MSFRossler}).}
\end{figure}

We study the width of the synchronized interval $l_{\varepsilon}$ as a function of number of edges. We start by considering a complete network, where there is an edge between each pair of the nodes. At each step, we remove an edge randomly keeping the network connected and compute the synchronized interval $l_{\varepsilon}$. We keep on removing edges until the network become minimally connected, further removal of edges will lead to disconnected network. We repeat this process several times. In Fig.~(\ref{lepsilon_eigrat_128_stddev}), we plot the average values of the synchronized interval $l_{\varepsilon}$ (solid red curve, left y-axis) and the eigenvalue ratio $\lambda_N/\lambda_2$ (dashed blue curve, right y-axis) for $x$-component coupled R\"ossler oscillators as functions of fraction of existing edges to the complete network $f=2E/N(N-1)$ for $N=128$ nodes, and $E$ is the number of existing edges. The solid and the dashed lines are the average values of $l_{\varepsilon}$ and $\lambda_N/\lambda_2$ respectively over 100 realizations and the shaded regions are the standard deviations.
For complete network $f=1$, $\lambda_1=0;\; \lambda_2=\ldots=\lambda_N=N$, and so $l_{\varepsilon}=\delta\alpha/N$ and $\lambda_N/\lambda_2 =1$. As we start removing edges from a complete network, the synchronized interval $l_{\varepsilon}$ first decreases a little and then increases when more edges are removed (inset of Fig.~(\ref{lepsilon_eigrat_128_stddev})). The reason of this small decrease in $l_{\varepsilon}$ is the presence of nodes which are connected to all other nodes. The largest eigenvalue $\lambda_N$ does not reduce from $\lambda_N=N$ till there exists a single node connected to all other nodes~\cite{Mohar1997}, but $\lambda_2$ decreases from $\lambda_2=N$ resulting a decrease in $l_{\varepsilon}$. When more edges are removed,  $\lambda_N$ reduces from the value $\lambda_N=N$ when there is no more nodes connected to all other the nodes and $l_{\varepsilon}$ start increasing. The synchronized interval $l_{\varepsilon}$ increases until it reaches a maximum value $(l_{\varepsilon}^{\rm{maximum}}\approx0.158)$ for the fraction of edges $f\approx0.082$. We denote this fraction of edges as the optimal fraction of edges $f^{\rm{optimal}}$ for which $l_{\varepsilon}$ is maximum. As further edges are removed, $l_{\varepsilon}$ decreases to zero, the network is no longer synchronizable. 
On the right y-axis of Fig.~(\ref{lepsilon_eigrat_128_stddev}), we plot the eigenvalue ratio $\lambda_N/\lambda_2$ as a function of $f$. The ratio increases monotonically increases as edges are removed and when the ratio increases beyond the ratio $\alpha_2/\alpha_1$ (dashed line in Fig~(\ref{lepsilon_eigrat_128_stddev})), the network lost its synchronizability and the synchronized interval becomes zero. The numerical results are averaged over 100 realizations, the shaded regions are the standard deviation of the adjacent curve. Thus, we observe that the synchronized interval increases (neglecting the small initial decrease) with random removal of edges and reaches a maximum value at $f^{\rm{optimal}}$ and then decreases.

\begin{figure}
\includegraphics[width=\columnwidth]{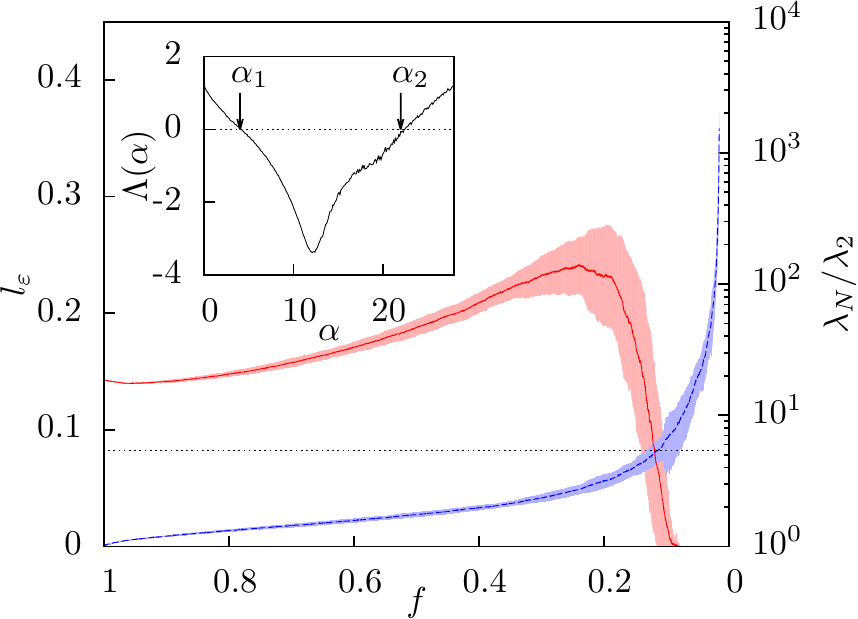}
\caption{\label{Lorenz_128N_MSF} The width of the synchronized interval $l_{\varepsilon}$ (solid red curve, left y-axis) is plotted as a function of fraction of edges $f$ for Lorenz oscillators with coupling from $y$-component to $x$-component of another oscillator. As edges are removed from undirected, unweighted complete networks, $l_{\varepsilon}$ increases and reaches a maximum value $l_{\varepsilon}^{\rm{maximum}}=0.241\pm0.021$  at $f^{\rm{optimal}}=$ and then decreases to zero. On the right y-axis we plot the eigenvalue ratio $\lambda_N/\lambda_2$ as a function of $f$. (Inset) The MSF is plotted for Lorenz oscillators with the same coupling function and here also the MSF is negative in a finite interval $(\alpha_1,\alpha_2)$ of parameter $\alpha$; here $\alpha_1 = 4.2$ and $\alpha_2=22.5$. The Lorenz parameters are $\sigma=10, \rho=28, \beta=2$. This is an averaged results over 100 realizations. The shaded regions are the standard deviations of the accompanying curves.}
\end{figure}

Similar result is also observed in coupled Lorenz oscillators with coupling from $y$-component to $x$-component of another oscillator. The dynamics of an isolated Lorenz oscillator is given by, $\dot{x}=\sigma(y-x),\dot{y}=x(\rho-z)-y,\dot{z}=xy-\beta z$, where $\sigma=10,\rho=28,\beta=2$~\cite{Lorenz1963}. Starting with a complete network, we remove edges randomly keeping the network connected and we compute the synchronized interval $l_{\varepsilon}$ as a function of fraction of edges $f$. We repeat the process several times. In Fig.~(\ref{Lorenz_128N_MSF}), we plot the average values of the synchronized interval $l_{\varepsilon}$ on left y-axis and the eigenvalue ratio $\lambda_N/\lambda_2$ on right y-axis, over 100 realizations, as functions of the fraction of edges $f$. The number of nodes is $N=128$. The shaded regions are the standard deviations. In the inset of Fig.~(\ref{Lorenz_128N_MSF}), we plot the MSF for Lorenz oscillator with coupling from $y$-component to $x$-component. The MSF is negative in a finite interval of parameter $\alpha$.  Form Fig.~(\ref{Lorenz_128N_MSF}), we find that the synchronized interval $l_{\varepsilon}$ increases as edges are removed from complete network, i.e. $f$ is decreased and at some value $f=f^{\rm{optimal}}$, the $l_{\varepsilon}$ becomes maximum and then decreases to zero as further edges are removed. The eigenvalue ratio $\lambda_N/\lambda_2$ increases monotonically with the decrease in $f$. This property of the synchronized intervals are observed in all coupled-oscillator systems for which the MSF is negative in a finite interval.

\begin{figure}
\includegraphics[]{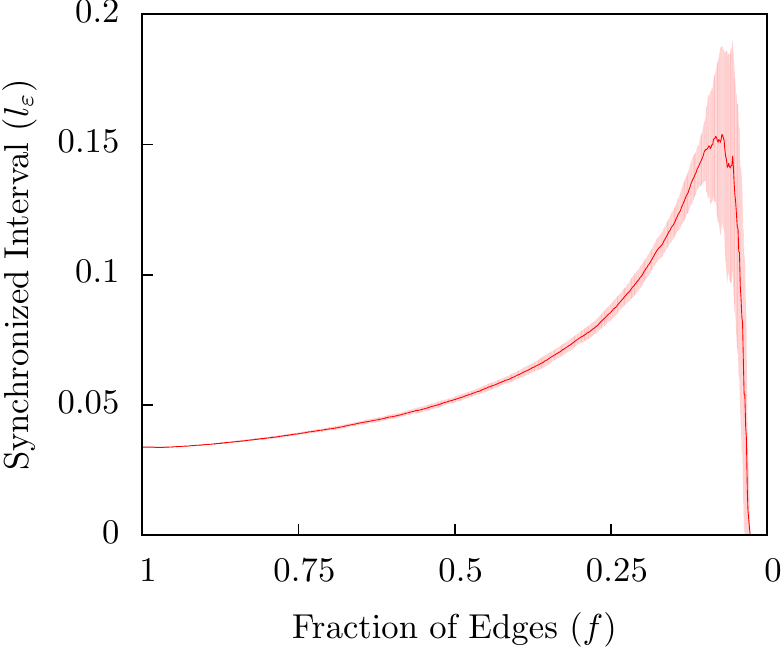}
\caption{\label{ER_lepsilon_eigrat_128_stddev} The average value of the synchronized interval $l_{\varepsilon}$ over 100 realizations (solid line) is plotted for Erd\"{o}s-R\'enyi random networks as a function of fraction of edges $f$. The number of nodes is 128. The dynamics of each node is given by identical R\"ossler oscillators with coupling in $x$-component (Fig.~\ref{MSFRossler}). The synchronized interval $l_{\varepsilon}$ is maximum at $f\approx0.08$. The shaded region represents the standard deviation for $l_{\varepsilon}$.}
\end{figure}

We have tested and observed similar dependence of the synchronized interval on the number of edges in Erd\"{o}s-R\'{e}nyi (ER) random networks~\ref{Erdos1960}. The ER random networks are generated by connecting two randomly selected nodes with a probability $p$. We tune this connection probability $p$ from zero to one. For very small $p$ the networks are not connected and we discard those networks, at $p=1$ the network is a complete network. At each value of $p$, we generate several ER random networks and compute the synchronized interval $l_{\varepsilon}$. In Fig.~\ref{ER_lepsilon_eigrat_128_stddev}, we plot the average value of $l_{\varepsilon}$ over 100 realizations (solid line) as a function of fraction of edges $f$ and the shaded region is the standard deviation of $l_{\varepsilon}$. The node dynamics are given by identical R\"ossler oscillators with $x$-component coupled and the R\"ossler parameters are $a=b=0.2$ and $c=7.0$ (Fig.~\ref{MSFRossler}). The average value of $l_{\varepsilon}$ is maximum at $f\approx0.08$ which matches with our previous observations on random networks generated by edge removal technique (Fig.~\ref{lepsilon_eigrat_128_stddev}).

Since we remove edges randomly from a complete network, the network eventually we acquire at the optimal value of the fraction of edges $f=f^{\rm{optimal}}$ is a random network. In random networks, the maximum and minimum nonzero eigenvalues of the coupling matrix $L$ are $\lambda_2 \approx Nf-\sqrt{2N\ln Nf(1-f)}$, and $\lambda_N \approx Nf+\sqrt{2N\ln Nf(1-f)}$~\cite{Mohar1997}. This approximation is better when the network size is large. Using the above forms of the eigenvalues in Eq.~(\ref{sync_interval2}), we write an analytic form for the width of the synchronized interval
\begin{equation}
l_{\varepsilon}^{\rm{analytic}} \approx \frac{\delta\alpha - (\alpha_1+\alpha_2)\sqrt{\frac{2(1-f)\ln N}{Nf}}}{Nf - 2\ln N(1-f)}.
\label{leps_analytical}
\end{equation}

\begin{figure}
\includegraphics[width=.9\columnwidth]{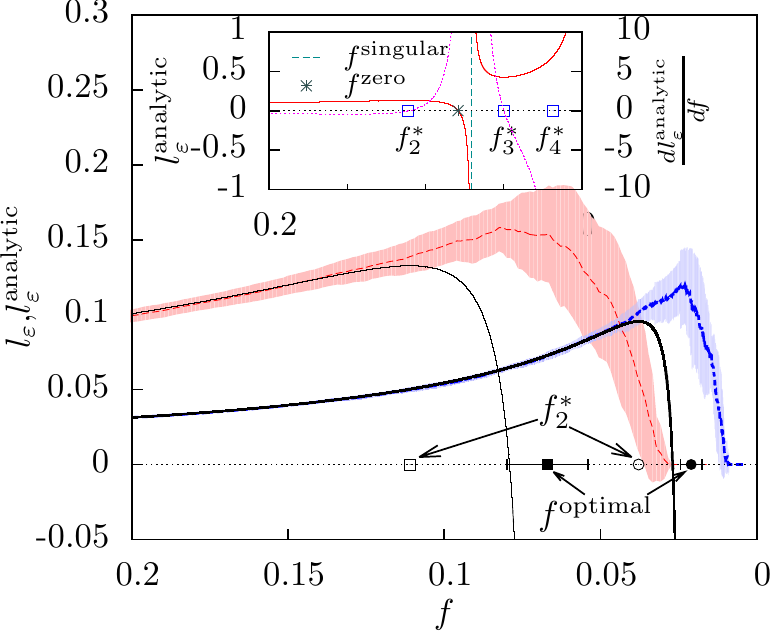}
\caption{\label{lepsilon_theory_numeric} We plot the numerical (dashed curves) value of width of synchronized interval $(l_{\varepsilon})$ and the analytical (solid curve) expression ($l_{\varepsilon}^{\rm{analytic}}$) from Eq.~(\ref{leps_analytical}) as a function of fraction of edges $f$ for $N=128$ nodes (thin curves) and $N=512$ nodes (thick curves) respectively. The open symbols (squares and circles) represent the second largest roots $f^*_2$ which correspond to the maximum of $l_{\varepsilon}^{\rm{analytic}}$ in Eq.~(\ref{leps_analytical}) (respectively for 128 and 512 nodes) and the solid symbols are the numerically obtained values of $f^{\rm{optimal}}$ at which the width of synchronized interval $l_{\varepsilon}$ is maximum.
(Inset) We plot $l_{\varepsilon}^{\rm{analytic}}$ from Eq.~(\ref{leps_analytical}) (solid line, left y-axis) and its derivative $d l_{\varepsilon}^{\rm{analytical}}(f)/d fs$ (dotted line, right y-axis) as functions of $f$ for $N=128$ nodes. The open squares are the three roots $(f^*_2,f^*_3,f^*_4)$ of the fourth order polynomial obtained from Eq.~(\ref{polynomial}). The singular point $f^{\rm{singular}}$ of $l_{\varepsilon}$ is represented by the dashed vertical line and the zero of $l_{\varepsilon}$, $f^{\rm{zero}}$ is shown by the asterisk. The dynamics of the coupled oscillators are given by R\"ossler oscillators and the parameters are same as in Fig.~(\ref{MSFRossler}).}
\end{figure}

In Fig.~(\ref{lepsilon_theory_numeric}), we plot the analytical expression of $l_{\varepsilon}^{\rm{analytic}}$ from Eq~(\ref{leps_analytical}) (solid lines) and the numerical value of the synchronized interval (dashed curves) for $N=128$ nodes (thin curves) and $N=512$ nodes (thick curves) respectively. 
For both network sizes, when the network is dense, i.e. $f$ is large both the analytical and numerical curves are in good agreement, but for small value of $f$ there is significant difference between these two curves which is due to the finite size of the networks. Here, we note that Eq.~(\ref{leps_analytical}) provide an approximation for the width of the synchronized interval and the approximation is better when the network size is large. As we can find from Fig.~(\ref{lepsilon_theory_numeric}), the difference between the analytical and numerical curves for 512 nodes is smaller than that of 128 nodes. The solid square and the solid circle correspond to the $f^{\rm{optimal}}$ values for which the width of the synchronized interval is maximum for 128 and 512 nodes respectively. The errorbars are the standard deviations. Though there is a deviation between the numerical and the analytical results, the qualitative nature of the analytical results are same as that of the numerical results, i.e. the analytic line $l_{\varepsilon}^{\rm{analytic}}$ increases as $f$ is reduced from one, then reaches a maximum at $f=f^*_2$ and then decreases to zero. We discuss details about $f^*_2$ in the next paragraphs.

From Eq.~(\ref{leps_analytical}), we find the local maxima and minima of $l_{\varepsilon}^{\rm{analytical}}$ by solving $d l_{\varepsilon}^{\rm{analytical}}(f)/d f = 0$ which lead to the following polynomial equation in $f$
\begin{eqnarray}
\frac{(\alpha_2+\alpha_1)}{\delta\alpha}\sqrt{\frac{\ln N}{2N}}\left[ f(3-2f) - \frac{2\ln N}{N+2\ln N}\right]
&&\nonumber \\
-\sqrt{f^3(1-f)}&=&0.
\label{polynomial}
\end{eqnarray}
We are interested in the roots (or zeros) of the Eq.~(\ref{polynomial}). Since Eq~(\ref{polynomial}) has fractional power in $f$, so we transform  it into the following fourth order polynomial equation in $f$ by taking square of both sides
\begin{eqnarray}
(4B+1)f^4 - (1+12B)f^3 + (4A+9)Bf^2 &&\nonumber \\
-6ABf + A^2B &=& 0,
\label{fourth_poly}
\end{eqnarray}
where, $A=2\ln N/(N+2\ln N)$ and $B=\ln N(\alpha_2+\alpha_1)^2/(2N\delta\alpha^2)$. Let $f^*_i;\;i=1,...,4$ be the four roots of the polynomial equation Eq.~(\ref{fourth_poly}) and we arrange these roots in the following descending order $f^*_1>f^*_2>f^*_3>f^*_4$. Here we do not provide the analytic expressions for $f^*_i;\;i=1,...,4$ due to their very complex structures~\cite{note_analytical}. Out of these four roots $f^*_1,f^*_2$ and $f^*_3$ correspond to the extrema of $l_{\varepsilon}^{\rm{analytical}}$, but $f^*_4$ does not. In the inset of Fig~(\ref{lepsilon_theory_numeric}), we plot the $l_{\varepsilon}^{\rm{analytic}}$ (solid line, left y-axis) and its derivative $d l_{\varepsilon}^{\rm{analytical}}(f)/d f$ (dotted line, right y-axis) as functions of $f$ when $N=128$. For clarity do not show $f^*_1$. Clearly, $f^*_2$ and $f^*_3$ correspond to $d l_{\varepsilon}^{\rm{analytical}}(f)/d f = 0$ while $f^*_4$ does not.

When $f$ is reduced from $f=1$, the first minimum of $l_{\varepsilon}^{\rm{analytical}}$ of Eq.~(\ref{leps_analytical}) happens at $f=f^*_1$, as we noted before that $l_{\varepsilon}$ decreases first and then increases as $f$ is decreased from $f=1$ (inset of Fig.~(\ref{lepsilon_eigrat_128_stddev})). Thus, the largest root $f^*_1$ occurs at a value of $f\sim 1$. When, $f$ is further reduced $l_{\varepsilon}^{\rm{analytical}}$ increases and reaches the maximum at the $f^*_2$, the second largest root of the fourth order polynomial equation Eq.~(\ref{fourth_poly}). The analytic expression for the width of synchronized interval in Eq.~(\ref{leps_analytical}) is zero at $f^{\rm{zero}} = 2\ln N (\alpha_2+\alpha_1)^2/[N(\alpha_2-\alpha_1)^2 + 2\ln N (\alpha_2+\alpha_1)^2]$, and it has a singular point $f^{\rm{singular}}=2\ln N/[N + 2\ln N]$~\cite{note_singular}, and clearly $f^{\rm{singular}} < f^{\rm{zero}}
$. The remaining two roots $f^*_3$ and $f^*_4$ occur at values  $f<f^{\rm{singular}}$.

\begin{figure}
\includegraphics[width=.9\columnwidth]{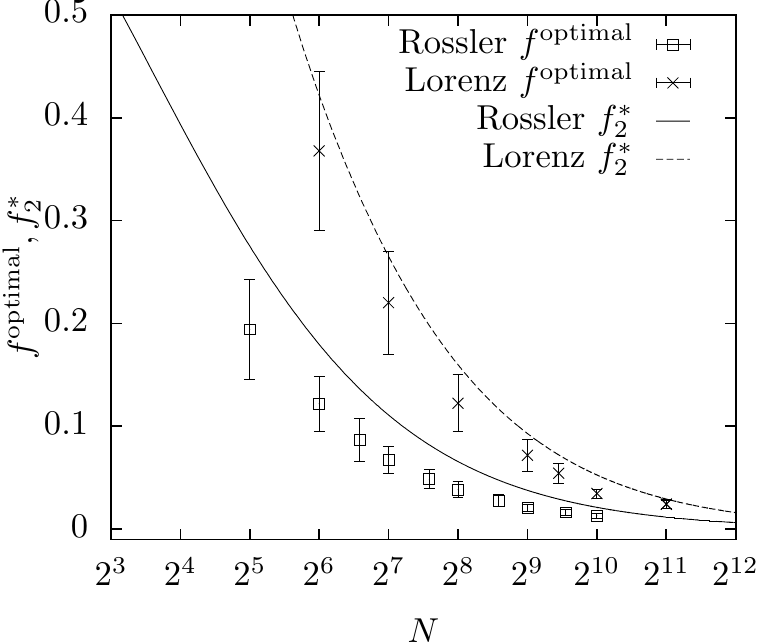}
\caption{\label{size_fmax_sigmafmax_ros_lor} The optimal value of fraction of edges $f^{\rm{optimal}}$ and the second largest root $f^*_2$ of the fourth order polynomial from Eq.~(\ref{polynomial}) are plotted as functions of number of nodes for R\"ossler oscillator (open square) and Lorenz oscillator (cross). The lines correspond to $f^*_2$ for R\"ossler (solid line) and Lorenz (dashed line) oscillators. Though the deviations of $f^{\rm{optimal}}$ and $f^*_2$ are large for small number of nodes, it reduces when the network size is large.}
\end{figure}

Thus, we are  interested in the second largest root $f^*_2$ at which the $l_{\varepsilon}^{\rm{analytic}}$ achieves its maximum value. In Fig.~(\ref{size_fmax_sigmafmax_ros_lor}), we plot the $f^{\rm{optimal}}$ (points with errorbar) and $f^*_2$ (lines) as functions of number of nodes for both R\"ossler and Lorenz oscillators. We can find that the agreement between the numerical results and the analytical results is good when the number of nodes is large. Thus, we conjecture that $f^*_2$ can fairly estimate the fraction of edges for which the produced undirected and unweighted random networks are synchronizable in wider interval of the coupling parameter.

\begin{figure}
\includegraphics[width=\columnwidth]{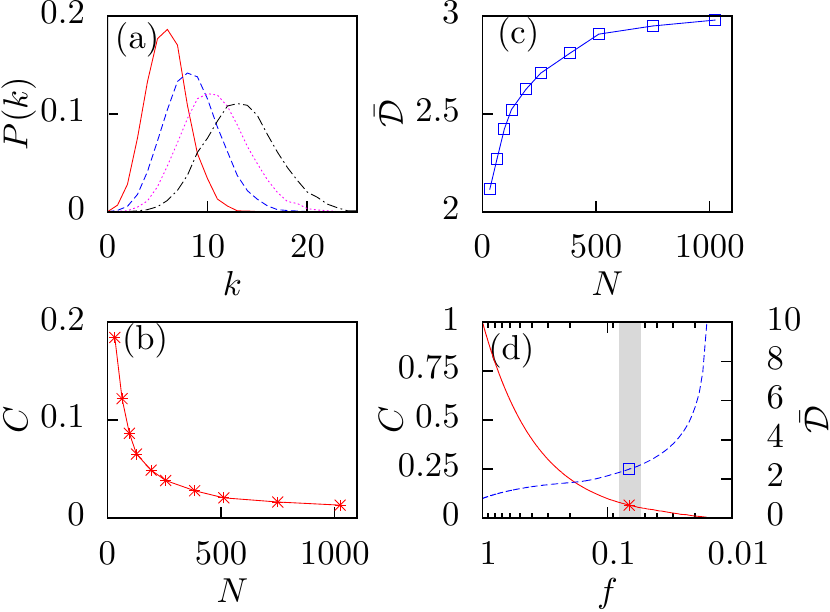}
\caption{\label{size_avgdist_cc_ros_128}(a) The degree distributions of the random networks at the optimal value of the fraction of edges $(f^{\rm{optimal}})$ are plotted. The number of nodes are respectively $N=32$ (red solid curve), $N=128$ (blue dashed curve), $N=512$ (pink dotted curve) and $N=1024$ (black dot-dashed curve). (a) The clustering coefficient $C$ at the $f^{\rm{optimal}}$ is plotted as a function of number of nodes $N$. (b) The average shortest path $\bar{\cal D}$ is plotted as a function of number of nodes. (c) Both the clustering coefficient $C$ (solid red line, left y-axis) and the average shortest path $\bar{\cal D}$ (dashed blue line, right y-axis) are plotted as functions of $f$ for a network with number of nodes $N=128$. The shaded region is the interval $(f^{\rm{optimal}}-\sigma_{f^{\rm{optimal}}},f^{\rm{optimal}}+\sigma_{f^{\rm{optimal}}})$ inside which the width of the synchronized interval achieves its maximum value. $\sigma_{f^{\rm{optimal}}}$ is the standard deviation of $f^{\rm{optimal}}$.}
\end{figure}

We study degree distributions $(P(k))$, average clustering coefficients $(C)$ and average shortest path lengths $({\cal D})$ to understand the structures of the networks, when the fraction of the number of edges to complete networks are given by the optimal value $f^{\rm{optimal}}$. Since we remove edges randomly from a complete network to get the optimal number of edges, the resultant networks are random networks. In Fig.~(\ref{size_avgdist_cc_ros_128})(a) we plot the degree distributions of the networks when the fraction of edges are give by $f^{\rm{optimal}}$. The number of nodes are respectively $N=32$ (red solid curve), $N=128$ (blue dashed curve), $N=512$ (pink dotted curve) and $N=1024$ (black dot-dashed curve). The degree distribution is Gaussian.

In Fig.~(\ref{size_avgdist_cc_ros_128})(b) we plot the average clustering coefficient $C=1/N\sum_i c_i$ of the networks at $f=f^{\rm{optimal}}$ as a function of number nodes $N$; where $c_i$ is the clustering coefficient of node $i$. The average clustering coefficient $C$ at $f^{\rm{optimal}}$ reduces with the increase in the number of nodes. In Fig.~(\ref{size_avgdist_cc_ros_128})(b), we plot the average shortest path length $\bar{\cal D}=1/N(N-1)\sum_{i,j} d_{ij}$ at $f=f^{\rm{optimal}}$ as a function of $N$; here $d_{ij}$ is the shortest path length between node $i$ and node $j$. Although the average shortest path length increases with the increase in network size, but the increment is small (in our case $3<\bar{\cal D}<2$). The average shortest path length at $f^{\rm{optimal}}$ varies as $\bar{\cal D}\sim\ln(N)$ which is a characteristic features for random network. 

In Fig.~(\ref{size_avgdist_cc_ros_128})(c), we plot the average clustering coefficient $C$ (solid red line, left y-axis) and the average shortest path length $\bar{\cal D}$ (dashed blue line, right y-axis) as a function of $f$ for the number of nodes $N=128$. In a complete network (i.e. $f=1$), obviously $C=1$ and $\bar{\cal D}=1$. As edges are removed, it can be easily anticipated that the average clustering coefficient $C$ reduces from $C=1$ and become zero when there is no more triangles in the network and the average shortest path length $\bar{\cal D}$ increases from $\bar{\cal D}=1$. The shaded area in Fig.~(\ref{size_avgdist_cc_ros_128})(c) indicate the interval $(f^{\rm{optimal}}-\sigma_{f^{\rm{optimal}}},f^{\rm{optimal}}+\sigma_{f^{\rm{optimal}}})$, inside which the width of synchronized interval achieves its maximum value, $\sigma_{f^{\rm{optimal}}}$ is the standard deviation of $f^{\rm{optimal}}$. Thus, the networks at $f=f^{\rm{optimal}}$ are random networks with fewer triangles and smaller average shortest path lengths~\cite{note_SW}. At this point, we want to emphasize that these networks are not optimized networks for synchronization. Instead in this article, we investigate the optimal value for number of edges for given the number of nodes, so that the generated random network have high chances to remain synchronizable in wider interval of coupling parameter.

To conclude, we have studied the statistical nature of the synchronized interval as a function of the number of edges in undirected, unweighted random networks. We have found the existence of optimal value for the number of edges, for which a random network will have better synchronization properties. We analytically derive the optimal value of number of edges for given number of network nodes, with known oscillator dynamics and coupling function and compare this results with the numerical results. In large networks, the analytical and the numerical results are in very good agreement. Thus, for given number of nodes, the oscillator dynamics and the coupling function, one is able to provide an estimation of the number of edges (using our analytic approach) for which the derived random networks will have better chance to remain synchronized in wider interval of coupling parameter. Synchronization in wider interval of coupling parameter is desirable in many real-world networks, such as power grid networks. Our approach can be extended with suitable modifications to provide an estimation of the transmission lines that could be used to connect the power generators and obtain maximum synchronization.

\section{\label{acknowledgements}Acknowledgements}

The author is thankful to SS Ray for many critical comments on the manuscript. The author acknowledges many useful discussions with  R. E. Amritkar and Amit Apte.


\begin{thebibliography}{99}
\bibitem{note_singular} Here we note that, although this is a singular point for $l_{\varepsilon}^{\rm{analytic}}$, but in practice one can obtain connected network with fraction of edges $f=f^{\rm{singular}}$ which is already reported earlier in Ref~\cite{BarahonaPRL2002}.
\bibitem{note_analytical} The analytical expressions for the roots of Eq.~(\ref{polynomial}) can be obtained using softwares like Mathematica,
\end{thebibliography}
\end{document}